\documentstyle[epsf,12pt]{article}
\newcommand{\simgt}{\lower.5ex\hbox{$\; \buildrel > \over \sim \;$}}
\newcommand{\simlt}{\lower.5ex\hbox{$\; \buildrel < \over \sim \;$}}
\begin{document}
\title{Earth effects on supernova neutrinos and their implications
for neutrino parameters}
\author{Keitaro Takahashi$^{\rm a}$ and Katsuhiko Sato$^{\rm a,b}$\\
{\it $^{\rm a}$Department of Physics, University of Tokyo,
7-3-1 Hongo, Bunkyo,}\\ {\it Tokyo 113-0033, Japan}\\
{\it $^{\rm b}$Research Center for the Early Universe, 
University of Tokyo,}\\
{\it 7-3-1 Hongo, Bunkyo, Tokyo 113-0033, Japan}}
\maketitle

\baselineskip=16pt

\begin{abstract}
We perform a detailed study of the Earth matter effects
on supernova neutrinos with neutrino oscillation parameter LMA and
small $\theta_{13}$. The Earth effects show significant dependences
on the neutrino path length inside the Earth and the value
of $\Delta m^{2}_{12}$. We investigate rather optimistically
a possibility that we can probe the value of $\Delta m^{2}_{12}$ 
by the Earth effects. We assume that $\theta_{12}$ and the
direction of the supernova are known with enough accuracy and that
the resonance that occurs at higher density in supernova envelope is 
completely nonadiabatic.
Further the neutrino spectra before neutrinos go through
the Earth are assumed to be known.
Then we show that making use of these dependences,
we can obtain implication for the value of $\Delta m^{2}_{12}$
by comparing the observed energy spectrum to the predicted one.
When SK detects neutrinos from supernova at 10kpc which traveled 
through the Earth (nadir angle $<$ 80 degree), $\Delta m^{2}_{12}$ can be
determined with an accuracy of $\sim 10\%$. 
In much of the neutrino-detection-time-$\Delta m^{2}_{12}$ plane, 
$\Delta m^{2}_{12}$ might be determined with an accuracy equal to or
better than $\pm 0.5 \times 10^{-5} {\rm eV}^{2}$.

\noindent
$PACS:$ 14.60.Pq; 14.60.Lm; 96.40.Tv; 97.60.Bw; 

\noindent
$Keywords:$ Neutrino oscillation; Supernovae; Earth effects;

\end{abstract}
\newpage

\section{Introduction}

A type II supernova is a prodigious source of neutrinos.
Almost all of the binding energy, $E_{b} \sim 10^{53}$erg,
is radiated away as neutrinos. These neutrinos carry information
about both the core collapse process and intrinsic properties of
the neutrinos. In fact, the detection of the neutrino burst from SN1987A
\cite{K2_SN1987A,IMB_SN1987A} induced a number of studies on these
areas \cite{Arafune,Smirnov1987A,Sato,Goldman}.

Neutrino oscillation, which is confirmed by the observations of
solar and atmospheric neutrinos, can affect the energy spectrum 
of supernova neutrino drastically.
Neutrinos of all flavors are produced in the high dense region of 
the iron core \cite{Totani} and 
interact with matter before emerging from the supernova. 
The presence of non-zero masses and mixing in vacuum among various 
neutrino flavors results in strong matter dependent effects, including 
conversion from one flavor to another. Hence, the observed neutrino flux
in the detectors may be dramatically different for certain neutrino flavors 
and for certain values of mixing parameters, due to neutrino oscillation.
Similar matter effect occurs in the Earth, too.

These oscillation effects on neutrino spectra depend strongly on
neutrino oscillation parameters: mixing angles and mass spectrum.
Some of them are firmly established but others are not.
All the existing experimental results on the 
atmospheric neutrinos can be well described in terms of the
$\nu_{\mu} \leftrightarrow \nu_{\tau}$ vacuum oscillation with
mass squared difference and the mixing angle given by \cite{Fukuda},
\begin{equation}
\Delta m^{2}_{\rm atm} \approx 7 \times 10^{-3} {\rm eV}^2 \; , \;   
\sin^{2} 2 \theta \approx 1 .
\end{equation}  
In contrast, for the observed $\nu_{e}$ suppression
of solar neutrinos four solutions are still
possible \cite{Bahcall,Krastev,Garcia}: 
large mixing angle (LMA), small mixing angle (SMA), 
low $\Delta m^{2}$ (LOW), and vacuum oscillation (VO).
\begin{eqnarray}
({\rm LMA}) \;\Delta m^{2}_{\odot} & \approx & (1 \sim 10) \times 10^{-5}  {\rm eV}^2 \; , \; 
 \sin^{2} 2 \theta_{\odot} \approx 0.7 \sim 0.95 \\
({\rm SMA}) \;\Delta m^{2}_{\odot} & \approx & (4 \sim 10) \times 10^{-6}  {\rm eV}^2 \; , \;
 \sin^{2} 2 \theta_{\odot} \approx (2 \sim 10) \times 10^{-3} \\
({\rm LOW}) \;\Delta m^{2}_{\odot} & \approx & (0.5 \sim 2) \times 10^{-7}  {\rm eV}^2 \; , \;
 \sin^{2} 2 \theta_{\odot} \approx 0.9 \sim 1.0 \\
({\rm VO}) \;\Delta m^{2}_{\odot} & \approx & (0.6 \sim 6) \times 10^{-10}  {\rm eV}^2 \; , \; 
 \sin^{2} 2 \theta_{\odot} \approx 0.8 \sim 1.0 
\end{eqnarray}
For $\theta_{13}$,
the mixing angle between mass eigenstate $\nu_{1}, \nu_{3}$,
only upper bound has been known from reactor experiment \cite{CHOOZ}
and combined three generation analysis \cite{Fogli,Yasuda,MinakataYasuda}.
Also the nature of neutrino mass hierarchy (normal or
inverted) is still a matter of controversy\cite{Minakata,Barger}.

Supernova is a completely different system from 
solar, atmospheric, accelerator, and reactor neutrinos 
in regard to neutrino energy and flavor of produced neutrinos, 
propagation length and so forth. Then neutrino emission 
from a supernova is expected to give valuable 
information that can not be obtained from neutrinos from
other sources.

There have been some studies on the supernova neutrino
as an oscillation parameter prober. Dighe and Smirnov \cite{Dighe}
have studied the role the supernova neutrinos can play in the
reconstruction of the neutrino mass spectrum.
Dutta et al. \cite{Dutta} showed numerically that the events involving oxygen
targets increase dramatically when there is neutrino mixing.
We have shown in our previous work \cite{KT2} that the degeneracy of 
the solutions of the solar
neutrino problem can be broken by the combination of the SK and SNO
detections of a future Galactic supernova.

The Earth effects have also been studied by several authors.
In our previous work \cite{KT1}, we analyzed numerically the time-integrated 
energy spectra in a simple case and found the possibility
to probe the mixing angle $\theta_{13}$.
Lunardini and Smirnov \cite{SmirnovEarth} performed a study
of the Earth effects taking the arrival directions of neutrinos into
account and showed that studies of the Earth effects
will select the solution of the solar neutrino problem, probe the
mixing $U_{e3}$ and identify the hierarchy of the neutrino mass spectrum.

In this paper, we perform a detailed study of the Earth matter effects
on supernova neutrinos and show that the detection of Earth matter effects
allows us to probe $\Delta m^{2}_{12}$ more accurately than by
solar neutrino observations.

\section{Three-flavor formulation}

In the framework of three-flavor neutrino oscillation, the time
evolution equation of the neutrino wave functions can be written as
follows:

\begin{equation}
i\frac{d}{dt}\left(
	\begin{array}{ccc}\nu_e\\ \nu_{\mu}\\ \nu_{\tau}
	\end{array}\right)
= H(t)\left(
	\begin{array}{ccc}\nu_e\\ \nu_{\mu}\\ \nu_{\tau}
	\end{array}\right) \label{Schrodinger}
\end{equation}
\begin{eqnarray}
H(t)  \equiv 
U\left(
	\begin{array}{ccc}
		0 & 0 & 0\\
		0 & \Delta m^2_{21} /2E & 0\\
		0 & 0 & \Delta m^2_{31} /2E
	\end{array}\right)U^{-1} 
   	+\left(
	\begin{array}{ccc}
		A(t) & 0 & 0\\
		0 & 0 & 0\\
		0 & 0 & 0
	\end{array}\right),
\end{eqnarray}
where $A(t)=\sqrt{2}G_{F}n_{e}(t)$, $G_{F}$ is Fermi constant, $n_{e}(t)$
is the electron number density, 
$\Delta m^2_{ij}$ is the mass squared differences, and $E$
is the neutrino energy. In case of antineutrino, the sign of 
$A(t)$ changes.
Here U is a unitary 3 $\times$ 3 mixing matrix in vacuum:
\begin{equation}
U  =  \left(\begin{array}{ccc}
c_{12}c_{13} & s_{12}c_{13} & s_{13}\\
-s_{12}c_{23}-c_{12}s_{23}s_{13} & c_{12}c_{23}-s_{12}s_{23}s_{13} 
& s_{23}c_{13}\\
s_{12}s_{23}-c_{12}c_{23}s_{13} & -c_{12}s_{23}-s_{12}c_{23}s_{13} 
& c_{23}c_{13}
\end{array}\right)\label{mixing_matrix},
\end{equation}
where $s_{ij} = \sin{\theta_{ij}}, c_{ij} = \cos{\theta_{ij}}$ 
for $i,j=1,2,3 (i<j)$.
We have here put the CP phase equal to zero in the CKM matrix.

In $H(t)$, the first term is the origin of vacuum oscillation, and
the second term A(t), which is the only time-dependent term in H(t), is
the origin of MSW effect.

\section{Determination of neutrino oscillation parameters}

In this section we summarize our previous studies on the effects
of neutrino oscillation on supernova neutrino and show that
future detection of neutrino from Galactic supernova allows us
to break the degeneracy of the solutions of the solar neutrino problem
and to probe $\theta_{13}$.

In \cite{KT2}, three-flavor neutrino oscillation in the star is studied.
We calculated the expected event rate and energy spectra, and their
time evolution at the SuperKamiokande(SK) and the Sudbury Neutrino
Observatory(SNO). For the calculations of them, we used a realistic
neutrino burst model based on numerical simulations of supernova explosions
\cite{Totani,Wilson} and employed a realistic density profile based on 
a presupernova model \cite{Woosley}. The distance between the supernova
and the Earth was set to 10kpc.

By solving numerically the differential equations (\ref{Schrodinger}) 
from the center of
supernova to the outside of supernova, we obtained conversion 
probabilities $P(\nu_{\alpha \rightarrow \beta})$, i.e., probabilities
that a neutrino of flavor $\alpha$ produced at the center of supernova
is observed as a neutrino of flavor $\beta$.

We assumed normal mass hierarchy
and used four sets of mixing parameters shown in table \ref{table:parameter}.
Here $\theta_{12}$ and $\Delta m_{12}^{2}$ correspond to the solutions of
solar neutrino problem and $\theta_{23}$ and $\Delta m_{13}^{2}$ correspond
to the solution of atmospheric neutrino problem. 
The value of $\theta_{13}$ is taken to
be consistent with current upper bound from reactor experiment \cite{CHOOZ}.
These models are named after their values of mixing angle:
LMA-L means that $\theta_{12}$ is set to be LMA of solar neutrino problem
and $\theta_{13}$ is large.

\begin{figure}
\begin{center}
\epsfxsize=3.5in
\epsffile{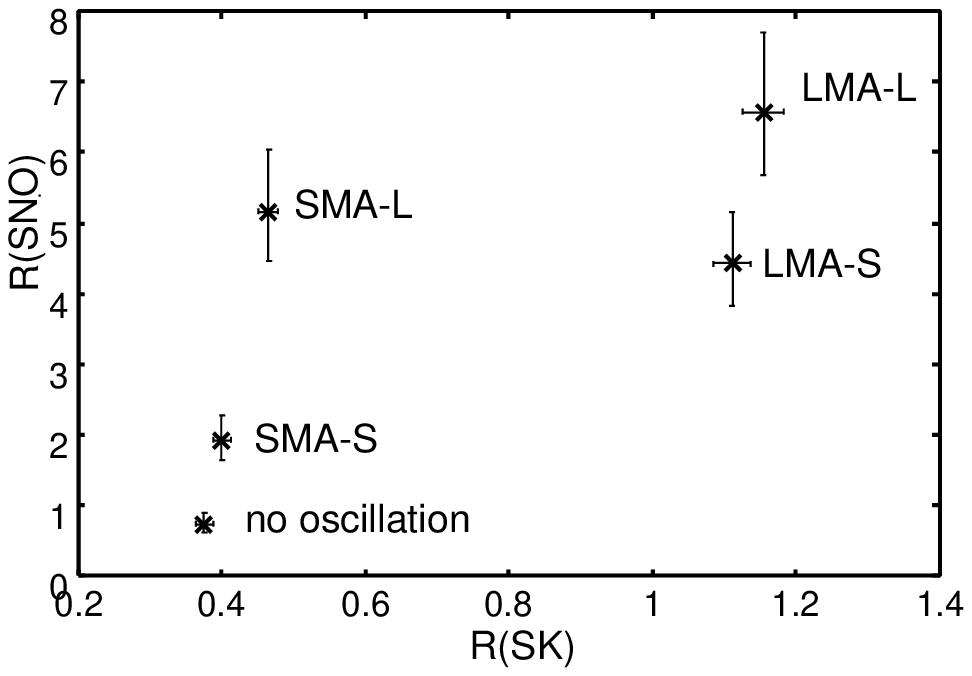}
\end{center}
\caption{ The plot of $R_{SK}$ vs. $R_{SNO}$ for all the models \cite{KT2}. 
The error-bars represent the statistical errors. 
 \label{figure:ratio}
}
\end{figure}

We then found that when there is neutrino oscillation, neutrino spectra are
harder than those in absence of neutrino oscillation. This is because
average energies of $\nu_{e}$ and $\bar{\nu}_{e}$ are smaller than those of
$\nu_{x}$(either of $\nu_{\mu}$, $\nu_{\tau}$, and their antineutrinos),
and neutrino oscillation produces high energy $\nu_{e}$ and $\bar{\nu}_{e}$
which was originally $\nu_{x}$. This feature was used as
a criterion of magnitude of neutrino oscillation.
We calculated the following ratio of events at both detectors:
\begin{equation}
R_{\rm SK} \equiv \frac{\makebox{number of events at } 30<E<70\rm{MeV}}
		{\makebox{number of events at } 5<E<20\rm{MeV}}
\end{equation}
\begin{equation}
R_{\rm SNO} \equiv \frac{\makebox{number of events at } 25<E<70\rm{MeV}}
		{\makebox{number of events at } 5<E<20\rm{MeV}}
\end{equation}
The plots of $R_{\rm SK}$ vs. $R_{\rm SNO}$ are shown in 
Fig.\ref{figure:ratio}. The errorbars include only statistical errors. 
The difference among the following three groups is clear:
(1)LMA-L and LMA-S, (2)SMA-L, and (3)SMA-S and no oscillation.
	
In \cite{KT1}, the Earth effects on the energy spectra are studied using
the result of \cite{KT2}.
For mass squared differences in table \ref{table:parameter}, 
the supernova neutrinos arrive at Earth in mass eigenstates.
This is because neutrino eigenstates with different masses
lose coherence on the way from the supernova to the Earth.
Since the eigenstates of Hamiltonian in matter differ from
the mass eigenstates in vacuum, supernova neutrinos begin to oscillate 
again in the Earth. The numbers of neutrinos of each mass eigenstates
at the surface of the Earth
are determined by neutrino oscillation in the supernova \cite{KT1}.

We analyzed numerically the time-integrated energy spectra of neutrino
in a mantle-core-mantle step function model of the Earth's matter
density profile. We assumed that neutrino arrived at the detectors
after traveling through the Earth along its diameter.
We then found that the Earth matter affects the
$\nu_{e}$ spectrum significantly in model LMA-S and $\bar{\nu}_{e}$ 
spectrum slightly in model LMA-S and LMA-L. In other models, 
there are no significant 
Earth effects. We concluded that we can
differentiate LMA-L from LMA-S, by observing the Earth matter effect
in $\nu_{e}$.

Thus we can distinguish all the four models in table \ref{table:parameter}.
But there are some ambiguities besides statistical errors.
One is the mass of the progenitor star. Supernovae with different progenitor
masses may result in different original neutrino spectra and neutrino
oscillation effects. Studies on this point are now in progress.
But dependence of shape of neutrino spectra on progenitor mass 
is not so large \cite{mass} and we would still be able to distinguish 
the models.
 
Supernova model which we use does not consider rotation of the
progenitor star. In general, however, stars rotate through their lives
and recent numerical simulation indicates that the rotation
facilitate supernova explosion \cite{Shimizu}.
Rotation of the progenitor star can affect its density profile 
and the dynamics of the neutrino oscillation in supernova.
	
\begin{figure}
\begin{center}
\epsfxsize=3.5in
\epsffile{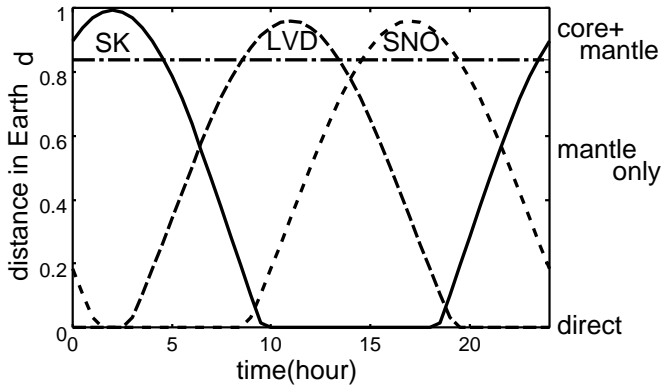}
\end{center}
\caption{ The distance which neutrinos travel in the Earth
as functions of the arrival time of neutrino burst \cite{SmirnovEarth}. 
We assume a supernova which located in the Galactic center. We fixed
$t=0$ as the time at which the star is aligned with the Greenwich
meridian.  \label{figure:distance}
}
\end{figure}

Another ambiguity is the direction of supernova.
The trajectories of neutrinos change according to the location of 
the supernova, the position of the detector and the time $t$ of the day
at which the burst arrives at the Earth \cite{SmirnovEarth}. 
Fig \ref{figure:distance} shows the dependence of the length $d$ of 
neutrino trajectory on the time $t$ for the three detectors,
SK, SNO and LVD. $d$ is in unit of the diameter of the Earth.
Neutrino trajectory can also be described by the nadir angle $\theta_{n}$
of the supernova with respect to the detector.
In the next section, we perform a detailed study of the Earth matter
effect.

\section{Detailed study of the Earth matter effect}

In this section, we concentrate on model LMA-S and
discuss the dependence of the Earth matter effect
on the distance which neutrinos travel in the Earth and $\Delta m^{2}_{12}$.
We consider three detectors: SK, SNO and LVD.
But energy spectra at SK and LVD in model LMA-L are the same
as in model LMA-S, since most of the events at SK and LVD
are induced by $\bar{\nu}_{e}$s which are influenced by the
Earth matter in both LMA-L and LMA-S. 
The dependence on $\Delta m^{2}_{12}$ is expected because the 
oscillation length in the Earth between mass eigenstates $\nu_{1}$ and
$\nu_{2}$ is comparable $(\sim 1000 {\rm km})$ to the diameter of the 
Earth in LMA-S model \cite{KT1}.

Here we use the realistic Earth density profile \cite{profile},
while we used step-like model in our previous work \cite{KT1}.
By solving (\ref{Schrodinger}) along the Earth density profile
numerically, we obtain conversion probability $P_{\nu_{i} \rightarrow
\nu_{\alpha}}$, i.e. probability that $i$th mass eigenstate at the
surface of the Earth is detected as neutrino of flavor $\alpha$.
Then combining these probabilities with the result of neutrino
oscillation in supernova \cite{KT2}, we obtain neutrino flux and
at each detector.

\subsection{neutrino events at SuperKamiokande}

SuperKamiokande (SK) is a water Cherenkov detector with 32,000 ton pure
water based at Kamioka in Japan. The relevant interactions of
neutrinos with water are as follows:
\begin{eqnarray}
\bar{\nu_e} +p & \rightarrow & n + e^+ \quad(\rm{CC}) 
\label{interaction:anuep}\\
\nu_e + e^- & \rightarrow & \nu_e + e^- \quad(\rm{CC \quad and \quad NC})\\
\bar{\nu_e} + e^- & \rightarrow & \bar{\nu_e} + e^- 
\quad(\rm{CC \quad and \quad NC})\\
\nu_x + e^- & \rightarrow & \nu_x + e^- \quad(\rm{NC})\\
\nu_e + {\rm O} & \rightarrow & {\rm F} + e^- \quad(\rm{CC})\\
\bar{\nu_e} + {\rm O} & \rightarrow & {\rm N} + e^+ 
\quad(\rm{CC})\label{interaction:nuebarO}
\end{eqnarray}
where CC and NC stand for charged current and neutral current
interactions, respectively. 
For the cross sections of these interactions, we refer to \cite{SKcross}.

SK is not available now due to the unfortunate accident \cite{SKaccident}
but is expected to be restored within the year. Since the number of
PMTs will decrease to half the original, energy threshold and 
energy resolution will become worse. The detection efficiency is expected
to be 90\% at 8MeV\cite{SKefficiency}. In the above interactions,
the $\bar{\nu_e}p$ CC interaction [Eq.(\ref{interaction:anuep})] has
the largest contribution to the detected events at SK. Hence the energy
spectrum detected at SK (including all the reactions) is almost the
same as the spectrum derived from the interaction Eq.(\ref{interaction:anuep}) 
only.

\subsection{neutrino events at SNO}

Sudbury Neutrino Observatory(SNO) is a water $\check{\rm{C}}$herenkov
detector based at Sudbury, Ontario. SNO is unique in its use of 1000
tons of heavy water, by which both the charged-current and
neutral-current interactions can be detected. The interactions of
neutrinos with heavy water are as follows,
\begin{eqnarray}
\nu_e + d & \rightarrow & p + p + e^-\quad(\rm{CC})\label{eq:SNO_CC_nue}\\
\bar{\nu_e} + d & \rightarrow & n + n + e^+\quad(\rm{CC})
	\label{eq:SNO_CC_anue}\\
\nu_x + d & \rightarrow & n + p + \nu_x\quad(\rm{NC})\\
\bar{\nu_x} + d & \rightarrow & n + p + \bar{\nu_x}\quad(\rm{NC})
\end{eqnarray}
The two interactions written in Eqs.(\ref{eq:SNO_CC_nue}) and
(\ref{eq:SNO_CC_anue}) are detected when electrons emit
$\check{\rm{C}}$herenkov light. These reactions produce electrons and
positrons whose energies sensitive to the neutrino energy, and hence the energy
spectra of electrons and positrons give us the information on the
original neutrino flux. In this work, we mainly take into account
these two charged current interactions, since
the number of neutral current events does not change by
neutrino oscillation. For the cross sections,
we refer to \cite{SNOcross}. The efficiency of detection is 
set to be one, because we have no information about it.

The SNO detector has also 7,000 tons of light water which can
be used to detect neutrinos. This can be considered to be a
miniature of SuperKamiokande (32,000 tons of light water).
Then the number of events detected by light water at SNO
is 7/32 of that at SuperKamiokande.

\subsection{neutrino events at LVD}

The Large Volume Detector (LVD) in the Gran Sasso Underground
Laboratory is a $\nu$ observatory mainly designed to study low
energy neutrinos from gravitational stellar collapse.
LVD consists of 2 kinds of detectors, namely: liquid scintillator,
for a total mass of 1840 tons, and streamer tubes for a total surface
of about $7000 {\rm m}^{2}$.

The bulk of events is due to the capture reaction:
\begin{equation}
\bar{\nu}_{e} + p \rightarrow n + e^{+} \; \; \; \; \; \; 
n + p \rightarrow d + \gamma_{2.2 {\rm MeV}}
\end{equation}
Further, about $5\%$ of the events are due to neutral current interactions
with ${}^{12}$C which deexcites emitting a 15.1 MeV $\gamma$.
Moreover, $3\%$ of events are due to elastic scattering of all neutrino
flavours on electrons, and less than $1\%$ to charged current 
interactions of $\nu_{e}$ and $\bar{\nu}_{e}$ with ${}^{12}$C nuclei.
For the cross sections, we refer to \cite{LVDcross}.
The appropriate detection efficiency curve is also taken into account
\cite{SmirnovEarth}.

\subsection{nadir angle dependence of energy spectra}

\begin{figure}
\begin{center}
\epsfxsize=4in
\epsffile{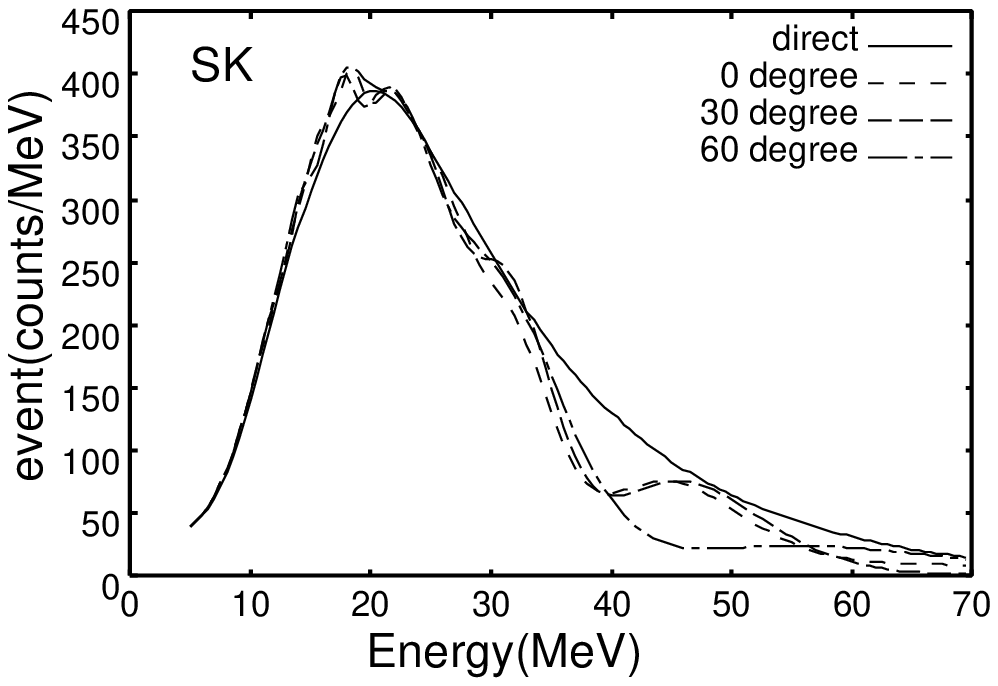}
\end{center}
\caption{ Nadir angle dependence of energy spectrum at SuperKamiokande.
Neutrino oscillation parameters are set to the values in model LMA-S
except $\Delta m^{2}_{12} = 2 \times 10^{-5} {\rm eV}^{2}$.
 \label{figure:SK_angle}
}
\end{figure}

\begin{figure}
\begin{center}
\epsfxsize=3.5in
\epsffile{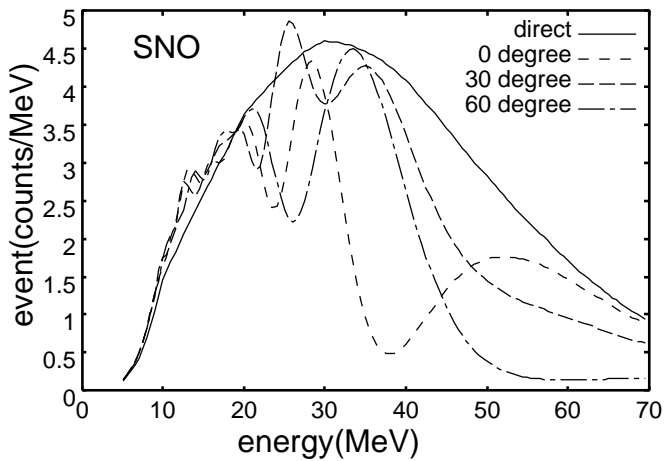}
\end{center}
\caption{  Nadir angle dependence of energy spectrum at SNO
taking only CC events into account.
Neutrino oscillation parameters are set to the values in model LMA-S
except $\Delta m^{2}_{12} = 2 \times 10^{-5} {\rm eV}^{2}$.
 \label{figure:SNO_angle}
}
\end{figure}

\begin{figure}
\begin{center}
\epsfxsize=3.5in
\epsffile{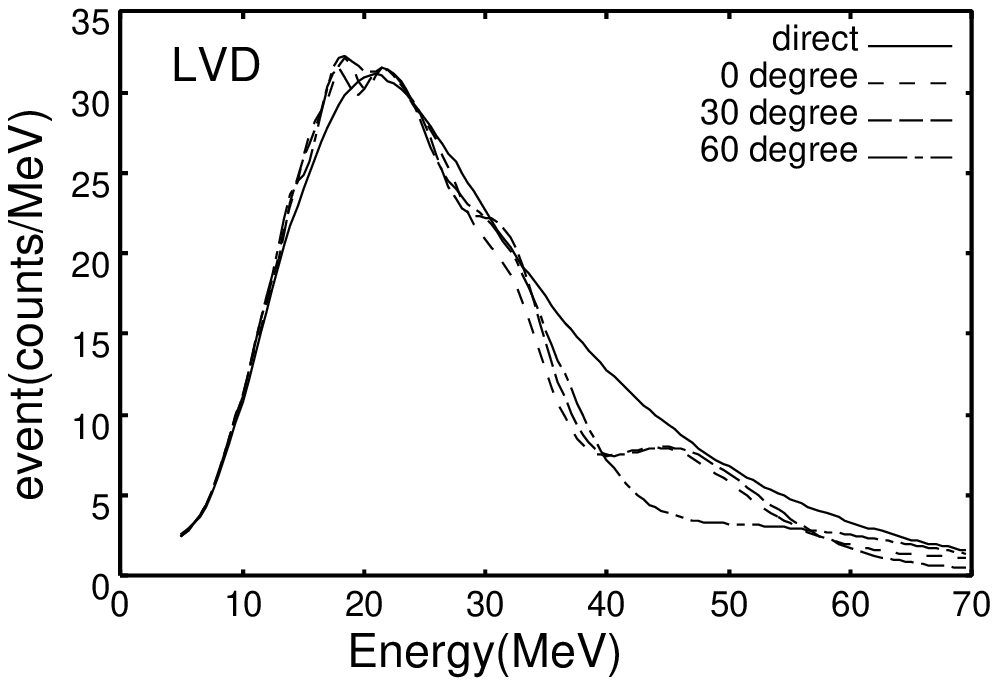}
\end{center}
\caption{  Nadir angle dependence of energy spectrum at LVD.
Neutrino oscillation parameters are set to the values in model LMA-S
except $\Delta m^{2}_{12} = 2 \times 10^{-5} {\rm eV}^{2}$.
 \label{figure:LVD_angle}
}
\end{figure}

Fig.\ref{figure:SK_angle}, \ref{figure:SNO_angle} and \ref{figure:LVD_angle} 
shows the nadir angle dependence of neutrino spectrum at SK, SNO and LVD, respectively. 
Neutrino oscillation parameters are set to the values in model LMA-S
except $\Delta m^{2}_{12} = 2 \times 10^{-5} {\rm eV}^{2}$.
Since the oscillation lengths are comparable to the radius of the Earth,
the Earth effects are strongly dependent on the nadir angle.

\section{$\Delta m^{2}_{12}$ dependence of energy spectra and
determination of $\Delta m^{2}_{12}$}

\subsection{method}

\begin{figure}
\begin{center}
\epsfxsize=4in
\epsffile{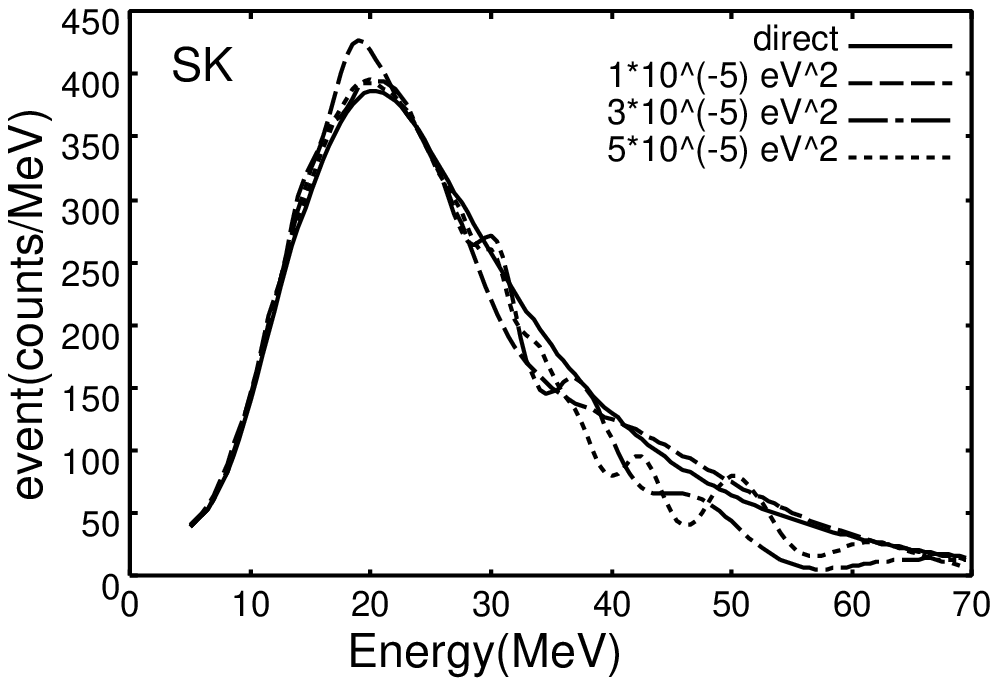}
\end{center}
\caption{  $\Delta m^{2}_{12}$ dependence of energy spectrum at
SuperKamiokande. Neutrino oscillation parameters are set to the 
values in model LMA-S except $\Delta m^{2}_{12}$. Nadir angle
is set to 0 degree.
 \label{figure:SK_mass}
}
\end{figure}

\begin{figure}
\begin{center}
\epsfxsize=3.5in
\epsffile{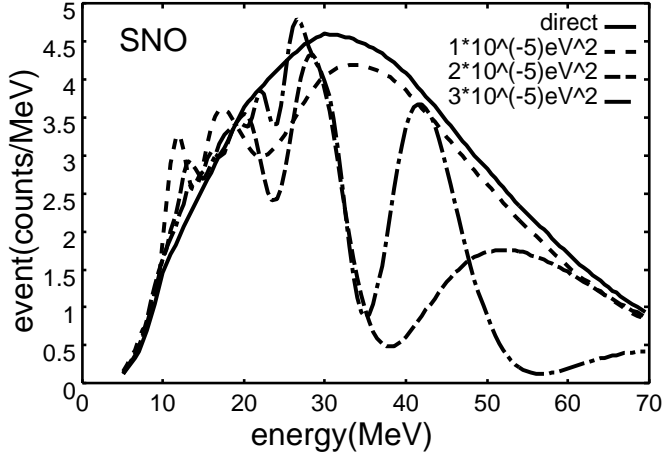}
\end{center}
\caption{  $\Delta m^{2}_{12}$ dependence of energy spectrum at
SNO taking only CC events into account. 
Neutrino oscillation parameters are set to the 
values in model LMA-S except $\Delta m^{2}_{12}$. Nadir angle
is set to 0 degree.
 \label{figure:SNO_mass}
}
\end{figure}

\begin{figure}
\begin{center}
\epsfxsize=3.5in
\epsffile{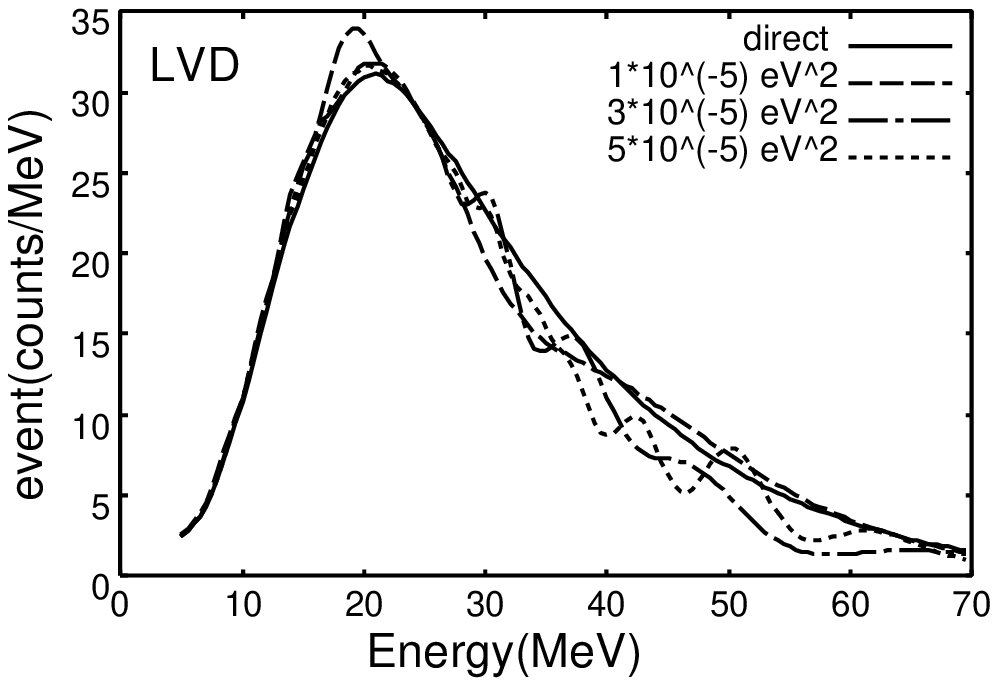}
\end{center}
\caption{  $\Delta m^{2}_{12}$ dependence of energy spectrum at
LVD. Neutrino oscillation parameters are set to the 
values in model LMA-S except $\Delta m^{2}_{12}$. Nadir angle
is set to 0 degree.
 \label{figure:LVD_mass}
}
\end{figure}

Fig.\ref{figure:SK_mass}, \ref{figure:SNO_mass} and \ref{figure:LVD_mass} 
shows the $\Delta m^{2}_{12}$ dependence of energy spectrum at SK, SNO and LVD. 
Neutrino oscillation parameters are set to the 
values in model LMA-S except $\Delta m^{2}_{12}$. Nadir angle
is set to 0 degree. The larger $\Delta m^{2}_{12}$ results in 
higher-frequency oscillation in the energy spectra with respect to
the energy. This is because the neutrino oscillation length is proportional
to the inverse of $\Delta m^{2}$.

As can be seen, there is 
significant dependence of the Earth effect on $\Delta m^{2}_{12}$.
Making use of this dependence, we can probe $\Delta m^{2}_{12}$
by the observations of the supernova neutrino spectra more accurately than
by the observations of solar neutrino,
if neutrino oscillation parameters are as in model LMA-S.

We assume that the direction of the supernova is given by direct
optical observations or by the experimental study of the neutrino 
scattering on electrons \cite{direction1,direction2}.
Other parameters of neutrino oscillation except $\Delta m^{2}_{12}$
are also assumed to be known with enough accuracy. 
But in fact, $\theta_{23}$ does not affect the Earth matter effects and
$\theta_{13}$ affects only the magnitude of the Earth matter effects
through the adiabaticity of the H-resonance in supernova envelope\cite{Dighe}.
If $\sin^{2}{\theta_{13}} \simlt 10^{-5}$, the H-resonance is completely
nonadiabatic for broad range of $\Delta m_{13}^{2}$
and in this case the Earth matter effects do not depend on
the value of $\theta_{13}$. So what we need with enough accuracy is
$\theta_{12}$ only if we assume the H-resonance is completely nonadiabatic.

\begin{figure}
\begin{center}
\epsfxsize=3.5in
\epsffile{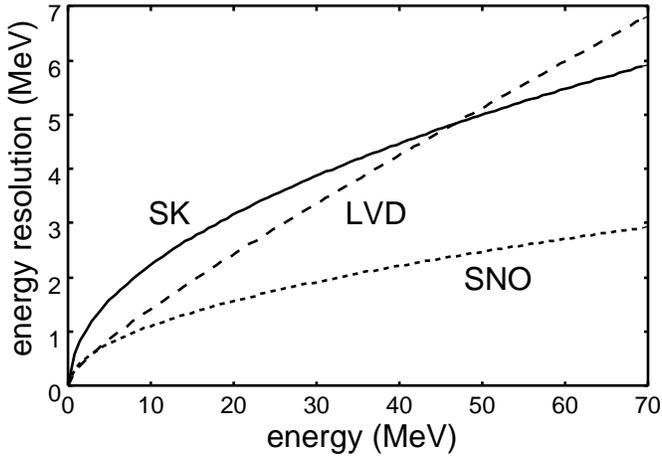}
\end{center}
\caption{  Energy resolution at each detector: SK (solid), SNO (dotted)
and LVD (dashed) \cite{SmirnovEarth}.
 \label{figure:resolution}
}
\end{figure}

Comparing the observed neutrino spectra to the predicted 
spectra of various values of $\Delta m^{2}_{12}$, we can determine 
the value of $\Delta m^{2}_{12}$.
We perform Monte Carlo simulation to obtain the expected observed spectra
at each detector. Here we take statistical fluctuation in number of neutrino 
events and energy resolution of each detector into account. 
Energy spectra are then binned according to the energy resolution.
For the value of the energy resolution, we refer to \cite{SmirnovEarth} but
that of SK after restoration is expected to be $\sqrt{2}$ times as bad as 
before the accident\cite{SKefficiency}.
The energy resolution of each detector is shown in Fig.\ref{figure:resolution}.

The simulated spectrum is compared to the predicted spectrum with
a $\chi^{2}$ method. 
The definition of the reduced $\chi^{2}$,
which is a function with respect to $\Delta m^{2}_{12}$, is,
\begin{equation}
\chi^{2}(\Delta m^{2}_{12}) = \frac{1}{d} \mathop{\sum}^{d}_{i=1} 
\frac{(N^{\rm sim}_{i}-N^{\rm pre}_{i}(\Delta m^{2}_{12}))^{2}}
{N^{\rm pre}_{i}(\Delta m^{2}_{12})},
\end{equation}
where $d$ is the number of bins and $N^{\rm sim}_{i}$ and 
$N^{\rm pre}_{i}(\Delta m^{2}_{12})$ are the simulated and predicted 
number of events in $i$th bin, respectively. The $\Delta m^{2}_{12}$ with 
the least $\chi^{2}(\Delta m^{2}_{12})$ is considered to be the value 
determined by this method.

\subsection{representative results}

\begin{figure}
\begin{center}
\epsfxsize=3.5in
\epsffile{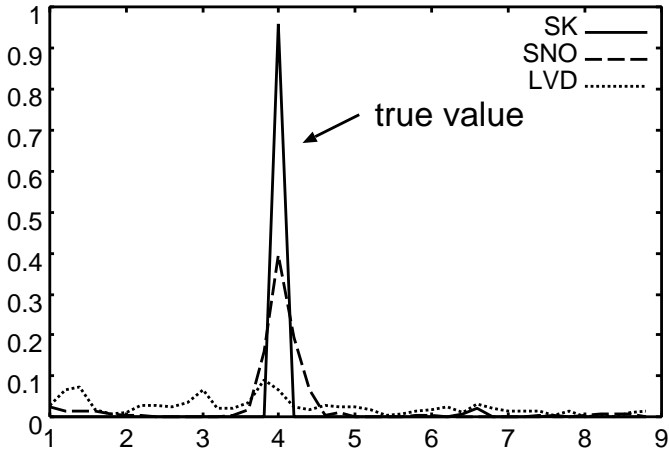}
\end{center}
\caption{  Relative frequencies that
the simulated spectrum based on the theoretical spectrum with
$\Delta m^{2}_{12} = 4 \times 10^{-5} {\rm eV}^{2}$ is identified
with the theoretical spectrum with the various values of $\Delta m^{2}_{12}$.
The nadir angle is set to be 0 degree at each detector 
and the $\chi^{2}$ method is performed with the data of each
detector only. 
 \label{figure:representative}
}
\end{figure}

We make 1000 ``supernovae'' and ``observe'' the neutrino spectra
assuming that the true value of $\Delta m^{2}_{12}$ is 
$4 \times 10^{-5} {\rm eV}^{2}$. Fig.\ref{figure:representative}
show the representative results of $\Delta m^{2}_{12}$ determination
by the method described above. These are relative frequencies that
the observed spectrum is identified
with the theoretical spectrum with the various values of $\Delta m^{2}_{12}$.
The nadir angle is set to be 0 degree at each detector 
and the $\chi^{2}$ method is performed with the data of each
detector only. 

As can be seen, if the supernova occur at the direction which
the nadir angle is 0 degree at SK or SNO, $\Delta m^{2}_{12}$
can be determined with a high accuracy $(4.00 \pm 0.07) \times 10^{-5} 
{\rm eV}^{2}$ or $(4.00 \pm 0.20) \times 10^{-5} {\rm eV}^{2}$
in case of SK and SNO, respectively. However, in case of LVD,
it is difficult to determine $\Delta m^{2}_{12}$. The difference in
the accuracy of determination of $\Delta m^{2}_{12}$ is due to the
difference in the properties of the detectors. 
Here the properties of the detectors mean the energy resolution,
the detectability of $\nu_{e}$ and the number of events.

Since the Earth effects are large in
$\nu_{e}$ channel and relatively small in $\bar{\nu}_{e}$ channel,
SK and LVD, at which most of events are $\bar{\nu}_{e}$, are disadvantageous. 
But the statistical error is very small at SK due to the large event number, 
$\sim 10000$ (10 kpc), and SK is the most efficient to identify the
Earth effects. SNO is also efficient due to its high energy-resolution 
and detectability of $\nu_{e}$ although the number of
events is small. Then LVD is not favorable for detecting the
Earth effects. The properties of the detectors are summarized in
the Table \ref{table:detector}.

It should be noted that the effectiveness of determining $\Delta m^{2}_{12}$
depend on the value of $\Delta m^{2}_{12}$ itself and nadir angle.
These dependences are discussed in the following subsection.

\subsection{determination of $\Delta m^{2}_{12}$}

\begin{figure}
\begin{center}
\epsfxsize=3.5in
\epsffile{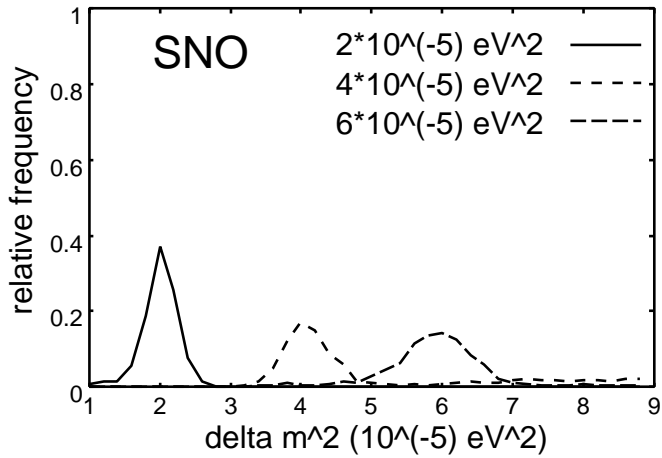}
\end{center}
\caption{  Relative frequencies at SNO that
the simulated spectrum based on the theoretical spectrum with
$\Delta m^{2}_{12} = 2 {\rm (solid)}, 4 {\rm (dashed)}, 6 {\rm (long-dashed)} 
\times 10^{-5} {\rm eV}^{2}$ is identified
with the theoretical spectrum with the various values of $\Delta m^{2}_{12}$.
The nadir angle is set to be 80 degree.
 \label{figure:SNO_best_mass}
}
\end{figure}

\begin{figure}
\begin{center}
\epsfxsize=3.5in
\epsffile{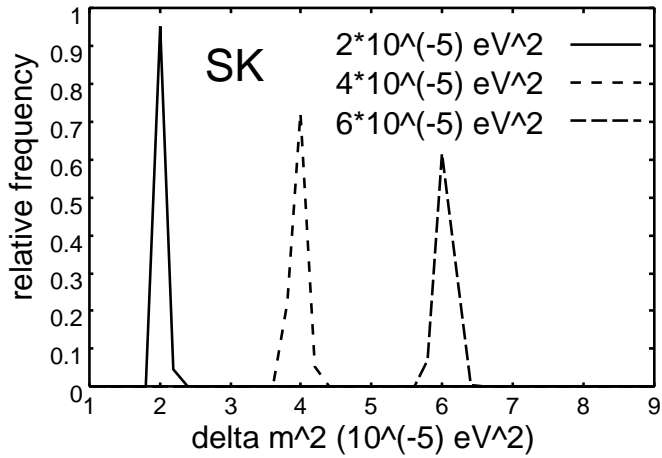}
\end{center}
\caption{  The same as in Fig.\ref{figure:SNO_best_mass} for SK.
 \label{figure:SK_best_mass}
}
\end{figure}

Fig.\ref{figure:SNO_best_mass} shows the relative frequencies at SNO that
the simulated spectrum based on the theoretical spectrum with
$\Delta m^{2}_{12} = 2 {\rm (solid)}, 4 {\rm (dashed)}, 6 {\rm (long-dashed)} 
\times 10^{-5} {\rm eV}^{2}$ is identified
with the theoretical spectrum with the various values of $\Delta m^{2}_{12}$.
The nadir angle is set to be 0 degree.
When $\Delta m^{2}_{12}$ is large, the oscillation in energy spectrum becomes
rapid and then statistical errors make it difficult to identify
the Earth effect and determine $\Delta m^{2}_{12}$.
However at SK, its large number of events allows us to determine
$\Delta m^{2}_{12}$ rather accurately irrespective of the true value of 
$\Delta m^{2}_{12}$ (Fig.\ref{figure:SK_best_mass}).

\begin{figure}
\begin{center}
\epsfxsize=3.5in
\epsffile{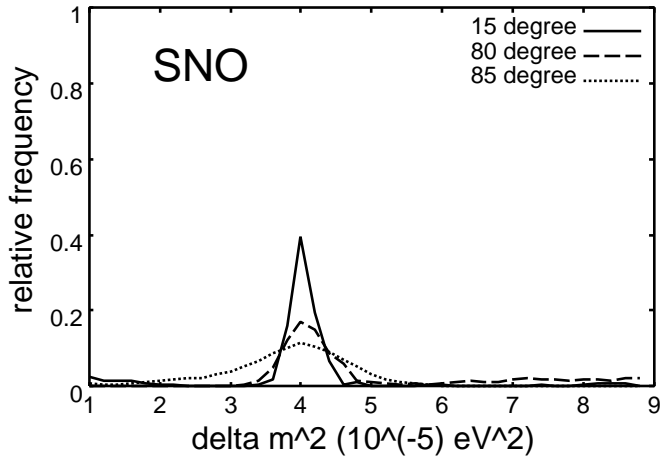}
\end{center}
\caption{  Relative frequencies at SNO at
various nadir angles: 15 (solid), 80 (dashed) and 85 (long-dashed). 
The value of $\Delta m^{2}_{12}$ is set to
be $4 \times 10^{-5} {\rm eV}^{2}$.
 \label{figure:SNO_best_angle}
}
\end{figure}

Fig.\ref{figure:SNO_best_angle} shows the relative frequencies at SNO at
various nadir angles: 15 (solid), 80 (dashed) and 85 (long-dashed). 
The value of $\Delta m^{2}_{12}$ is set to
be $4 \times 10^{-5} {\rm eV}^{2}$.
Since the oscillation length of 60 MeV $\nu$ 
in Earth matter is $\sim 6000$km \cite{KT1}, the Earth effects
do not appear unless nadir angle is smaller than $\sim 80$ degree.

Now we consider the problem, 'How accurately can we determine
$\Delta m^{2}_{12}$ when the true value of $\Delta m^{2}_{12}$
and the arrival time of neutrinos are given?'

As is discussed above, when the nadir angle at SK
is smaller than 80 degree (time $= 0 \sim 8$ and $19 \sim 24$ hour), 
the value of $\Delta m^{2}_{12}$ can be determined accurately
using only the SK data
irrespective of the value of $\Delta m^{2}_{12}$ in the allowed
region of the solar neutrino problem.
When the nadir angle is smaller than 80 degree at SNO and greater
than 80 degree at SK (time $= 10 \sim 19$ hour), 
the value of $\Delta m^{2}_{12}$ can be 
determined accurately using only the SNO data if the true value of
$\Delta m^{2}_{12}$ is smaller than $\sim 4 \times 10^{-5} {\rm eV}^{2}$.
However, if $\Delta m^{2}_{12} > 4 \times 10^{-5} {\rm eV}^{2}$,
the single data of SNO is not sufficient.
We performed then a $\chi^{2}$ method
combining data of various detectors for time $= 8 \sim 19$ hour.

\begin{figure}
\begin{center}
\epsfxsize=4in
\epsffile{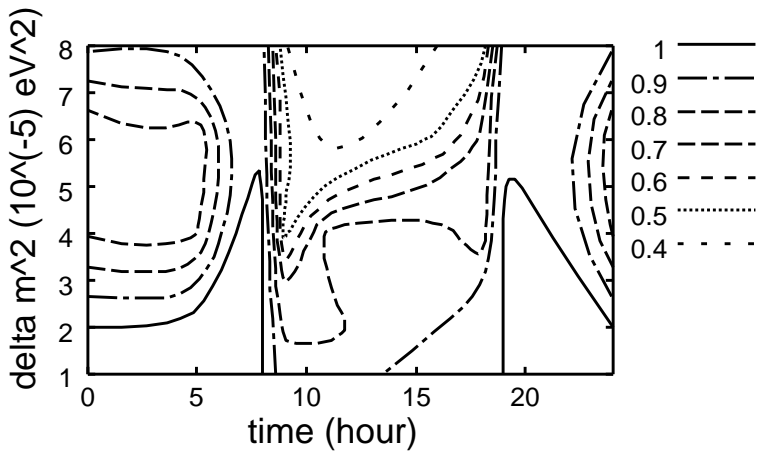}
\end{center}
\caption{ Contour map of the probability
that the value $\Delta m^{2}_{12}$ can be determined with an accuracy
equal to or better than $\pm 0.5 \times 10^{-5} {\rm eV}^{2}$.
 \label{figure:contour}
}
\end{figure}

Fig.\ref{figure:contour} is the contour map of the probability
that the value $\Delta m^{2}_{12}$ can be determined with an accuracy
equal to or better than $\pm 0.5 \times 10^{-5} {\rm eV}^{2}$.
The probability is more than 0.7 for three fifths of the 
time-$\Delta m^{2}_{12}$ plane.

\section{Discussion and conclusion}

In this paper, we performed a detailed study of the Earth matter effects
on supernova neutrinos with neutrino oscillation parameter LMA and
small $\theta_{13}$. We showed that we can probe $\Delta m^{2}_{12}$
accurately by comparing the observed energy spectra to the predicted one.
In much of the time-$\Delta m^{2}_{12}$ plane, $\Delta m^{2}_{12}$ can be 
determined with an accuracy equal to or better than 
$\pm 0.5 \times 10^{-5} {\rm eV}^{2}$. 
When SK detect neutrinos from supernova at 10kpc which traveled 
through the Earth (nadir angle $<$ 80 degree), $\Delta m^{2}_{12}$ can be
determined with an accuracy of $\sim 10\%$. This accuracy is amazing
compared to that from current solar neutrino experiments, which determine
only the order of $\Delta m^{2}_{12}$. But as we stated before, $\theta_{12}$ 
and the direction of the supernova
need to be known with enough accuracy. How much our results are affected
by the uncertainties of the other parameters will be our future work.

$\Delta m^{2}_{12}$ can also be probed by KamLAND experiment\cite{KamLAND}.
If the solution of solar neutrino problem is LMA,
$\Delta m^{2}_{12}$ is expected to be known by three-year
data-taking as precisely as by supernova neutrino discussed above.
There is some possibility that 
we can know it from a Galactic supernova before the completion
of the KamLAND experiment. Even if the results of KamLAND experiment
is earlier than future galactic supernova, it is important to
cross-check the results of ground-based experiments
by astrophysical observations. Furthermore, it is meaningful
itself to notice that we can obtain implication for $\Delta m^{2}$
from supernova neutrino.

We assumed in this paper that the neutrino spectra at the surface of the
Earth were given. But since the original fluxes based on the numerical 
supernova model have some ambiguities, the sensitivity to $\Delta m_{12}^{2}$
might be worse than our estimation. Estimation of uncertainties of
the numerical supernova model is hard because there is currently
no successful simulation other than that by the Lawrence Livermore group
\cite{Wilson}. There are, however, some studies about the temperatures 
of the produced neutrinos. According to them, they are typically\cite{Janka},
\begin{equation}
\langle E_{\nu_{e}} \rangle = 10 - 12 {\rm MeV}, \;\;
\langle E_{\bar{\nu}_{e}} \rangle = 14 - 17 {\rm MeV}, \;\;
\langle E_{\nu_{x}} \rangle = 24 - 27 {\rm MeV}.
\end{equation}
On the other hand, in the model that we used, they are
\begin{equation}
\langle E_{\nu_{e}} \rangle \sim 13 {\rm MeV}, \;\;
\langle E_{\bar{\nu}_{e}} \rangle \sim 16 {\rm MeV}, \;\;
\langle E_{\nu_{x}} \rangle \sim 23 {\rm MeV}, \;\;
\end{equation}
although the spectra are not exactly the Fermi-Dirac distribution\cite{Totani}.
These values seem close to the typical values and the analysis taking these
uncertainties into account will be performed in our future work.

Furthermore, supernova neutrino spectra depend on the mass of 
the progenitor star. However, we will be able to reconstruct the neutrino 
fluxes at the surface of the Earth from the data from the detector
which detect the neutrinos directly from the supernova.
Although the statistical errors then become larger and the accuracy of
the determination of $\Delta m^{2}_{12}$ becomes worse,
the Earth effects on the supernova neutrinos will still give
valuable information which cannot be obtained from the other
neutrino sources.

\section{Acknowledgments}

We would like to thank A.Suzuki for advice on cross sections of neutrino
interaction in scintillator experiment. 
We also thank Super Kamiokande people including Y. Totsuka, Y. Suzuki,
T.Kajita, M. Nakahata and Y. Fukuda for showing the most recent SK data, and
fruitful discussion.
This work was supported in part
by Grants-in-Aid for Scientific Research provided by the Ministry
of Education, Science and Culture of Japan through Research Grant
No.07CE2002.

\clearpage

\clearpage

\begin{table*}
\caption{Sets of mixing parameter for calculation
	\label{table:parameter}}
\begin{center}
\begin{tabular}{ccccccc}
model  & $\sin^{2} 2 \theta_{12}$ & $\sin^{2} 2 \theta_{23}$ & $\sin^{2} 2 \theta_{13}$ 
& $\Delta m_{12}^{2}({\rm eV}^{2})$  & $\Delta m_{13}^{2}({\rm eV}^{2})$ 
& $\nu_{\odot}$ problem  \\ \hline 
LMA-L &  0.87  & 1.0 & 0.043 & $7.0 \times 10^{-5}$ & $3.2 \times 10^{-3}$ & LMA \\ 
LMA-S &  0.87  & 1.0 & $1.0 \times 10^{-6}$ & $7.0 \times 10^{-5}$ & $3.2 \times 10^{-3}$ 
& LMA \\   
SMA-L &  $5.0 \times 10^{-3}$  & 1.0 & 0.043  & $6.0 \times 10^{-6}$ & $3.2 \times 10^{-3}$ 
& SMA \\ 
SMA-S &  $5.0 \times 10^{-3}$  & 1.0 & $1.0 \times 10^{-6}$ & $6.0 \times 10^{-6}$ & $3.2 \times 10^{-3}$
& SMA \\ 
\end{tabular}
\end{center}
\end{table*}

\begin{table*}
\caption{Properties of the detectors \label{table:detector}}
\begin{center}
\begin{tabular}{cccc} 
detector       & SNO &    SK   &  LVD  \\ \hline
main event & ${\nu}_{e},\bar{\nu}_{e}$ & $\bar{\nu}_{e}$ & $\bar{\nu}_{e}$  \\
number of events (10{\rm kpc}) & 300 & 10000 & 800    \\ 
energy resolution & $<$ 3 MeV & $<$ 6 MeV & $<$ 7 MeV \\ 
\end{tabular}
\end{center}
\end{table*}


\begin{thebibliography}{99}

\bibitem{K2_SN1987A} K. Hirata et al.,
	{Phys. Rev. Lett. {\bf 58}, 1490 (1987)}.

\bibitem{IMB_SN1987A} R. M. Bionta et al.,
	{Phys. Rev. Lett. {\bf 58}, 1494 (1987)}.

\bibitem{Arafune} J.Arafune and M.Fukugita, 
        {Phys. Rev. Lett. {\bf59}, 367 (1987)}.

\bibitem{Smirnov1987A}C. Lunardini, A. Yu. Smirnov, {hep-ph/0009356}.

\bibitem{Sato}K.Sato and H.Suzuki, 
        {Phys. Rev. Lett. {\bf 58}, 2722 (1987)}.

\bibitem{Goldman}I.Goldman et al., 
        {Phys. Rev. Lett. {\bf 60}, 1789 (1988)}.

\bibitem{Totani}
T.Totani, K.Sato, H.E.Dalhed and J.R.Wilson,
Astrophys. J. 496 (1998) 216.

\bibitem{Fukuda}
Y. Fukuda et al., Phys. Rev. Lett. 82 (1999) 2644.

\bibitem{Bahcall}
J.N.Bahcall, P.I.Krastev and A.Y.Smirnov,
Phys. Rev. D 58 (1998) 096016.  

\bibitem{Krastev}
P.I.Krastev, hep-ph/9905458. 

\bibitem{Garcia}
M.C.Gonzalez-Garcia, P.C.de Holanda, C.Pena-Garay and J.W.F.Valle,
hep-ph/9906469.

\bibitem{CHOOZ}M.Apollonio et al.,{Phys. Lett B {\bf 466}, 415 (1999)}.

\bibitem{Fogli}G. L. Fogli et al. ,{hep-ph/0104221}.

\bibitem{Yasuda}
O.Yasuda and H. Minakata, {hep-ph/9602386}

\bibitem{MinakataYasuda}
H.Minakata and O.Yasuda, {Phys. Rev. D {\bf 56}, 1692 (1997)}.

\bibitem{Minakata}H. Minakata and H. Nunokawa, {hep-ph/0010240}.

\bibitem{Barger}
V. Barger, D. Marfatia and B. P. Wood, hep-ph/0202158.

\bibitem{Dighe}
A. S.Dighe and A. Y. Smirnov, 
Phys. Rev. D 62 (2000) 033007. 

\bibitem{Dutta}
G. Dutta, D.Indumathi, M.V.N.Murthy and G.Rajasekaran, 
Phys. Rev. D 61 (2000) 013009.

\bibitem{KT2}K.Takahashi, M.Watanabe, K.Sato and T.Totani, 
	{Phys. Rev. D accepted}, hep-ph/0105204.

\bibitem{KT1}K.Takahashi, M.Watanabe and K.Sato, 
	{Phys. Lett. B {\bf 510}, 189 (2001)}.

\bibitem{SmirnovEarth} C.Lunardini and A.Y.Smirnov,
{hep-ph/0106149}.

\bibitem{Wilson}J. R. Wilson, R. Mayle, S. Woosley, T. Weaver,
	{Ann. NY Acad. Sci. {\bf 470}, 267 (1986)}. 

\bibitem{Woosley}S. E. Woosley and T. A. Weaver, 
	{ApJ. Suppl. {\bf 101}, 181 (1995)}.

\bibitem{mass} R.Mayle, Ph.D.Thesis, University of California (1987);
               R.Mayle, J.R.Wilson, and D.N.Schramm, 
               {Astrophys. J. {\bf 318}, 288 (1987)}.

\bibitem{Shimizu}
T. M. Shimizu, T. Ebisuzaki, K. Sato, and S. Yamada,
{Astrophys. J. {\bf 553}, 756 (2001)}.

\bibitem{profile} A.M.Dziewonski and D.L.Anderson,
Phys.Earth.Planet.Inter. {\bf 25}(1981)297.

\bibitem{SKcross} Y.Totsuka, {Rep. Prog. Phys. {\bf 55}, 377 (1992)};
                  K.Nakamura, T.Kajita and A.Suzuki, {\em Kamiokande},
in {\em Physics and Astrophysics of Neutrino}, edited by M.Fukugita 
and A.Suzuki (Springer-Verlag, Tokyo, 1994).

\bibitem{SKaccident}
See, http://www-sk.icrr.u-tokyo.ac.jp/index.html

\bibitem{SKefficiency} Y.Totsuka, private communication.

\bibitem{SNOcross} S.Ying, W.C.Haxton and E.M.Henley, 
        {Phys. Rev. D {\bf 40}, 3211 (1989)}.

\bibitem{LVDcross} M.Fukugita, Y.Kohyama and K.Kubodera,
IASSNS-AST 88/25.

\bibitem{direction1} J.F.Beacom and P.Vogel,
Phys.Rev. {\bf D 60} (1999) 033007.

\bibitem{direction2} S.Ando and K.Sato, hep-ph/0110187.

\bibitem{KamLAND}
H. Murayama and A. Pierce,
Phys. Rev. {\bf D 65} (2002) 013012. 

\bibitem{Janka}
H.-T.Janka, in: F. Giovannelli and G. Mannocchi (eds.),
Proc. Vulcano Workshop 1992 {\it Frontier Objects in Astrophysics
and Particle Physics}, Conf. Proc. Vol. 40 (Soc. Ital. Fis).

\end{thebibliography}
\end{document}